# Proxy dynamic delegation in grid gateway

**Federico Calzolari**[1]
*Scuola Normale Superiore*
*Piazza Cavalieri 7, 56126 Pisa PI, ITALY*
*E-mail:* `federico.calzolari@sns.it`

**Daniele Licari**
*Scuola Normale Superiore*
*Piazza Cavalieri 7, 56126 Pisa PI, ITALY*
*E-mail:* `daniele.licari@sns.it`

Nowadays one of the main obstacles the research comes up against is the difficulty in accessing the required computational resources. Grid is able to offer the user a wide set of resources, even if they are often too hard to exploit for non expert end user. Use simplification has today become a common practice in the access and utilization of Cloud, Grid, and data center resources. With the launch of L-GRID gateway, we introduced a new way to deal with Grid portals. L-GRID is an extremely light portal developed in order to access the EGI Grid infrastructure via Web, allowing users to submit their jobs from whatever Web browser in a few minutes, without any knowledge about the underlying Grid infrastructure.

---

[1]  Speaker

## 1. Introduction

While most of the research, both in science and humanities, requires a growing computational power and data storage, the access to the Grid computing resources is nowadays too complex for a non-practiced researcher. Grid is today a very powerful tool for a few people, usually involved in big projects in physics, computational chemistry, and biomedicine. Use simplification has today become a common practice in the access and utilization of Cloud, Grid, and data center resources [1] [2].

Our idea is to give users a very simple tool to access the European Grid Infrastructure EGI resources [3], without requiring users to be expert in computer science or distributed computing architecture. The original idea has been to split the certificate management between client and server. This entails a simplification in the client side software implementation, committing the best part of the procedure to the server.

## 2. Aims

With the launch of L-GRID gateway, we introduced a new way to deal with Grid portals. L-GRID is an extremely light portal developed in order to access the EGI Grid infrastructure via Web, allowing users to submit their jobs from whatever Web browser in a few minutes, without any knowledge about the underlying Grid infrastructure.

## 3. The Grid portal architecture

The whole architecture, implemented as client-server architecture, is based on the Globus gLite Grid middleware and Java Commodity Grid Kits (CoG) library. The client side application is based on a java applet, running both on Windows, Linux and Mac operating systems; it only needs a Web browser connected to the Internet. The server relies on a Globus - gLite User Interface with a Web portal provided by an Apache/Tomcat server [4].

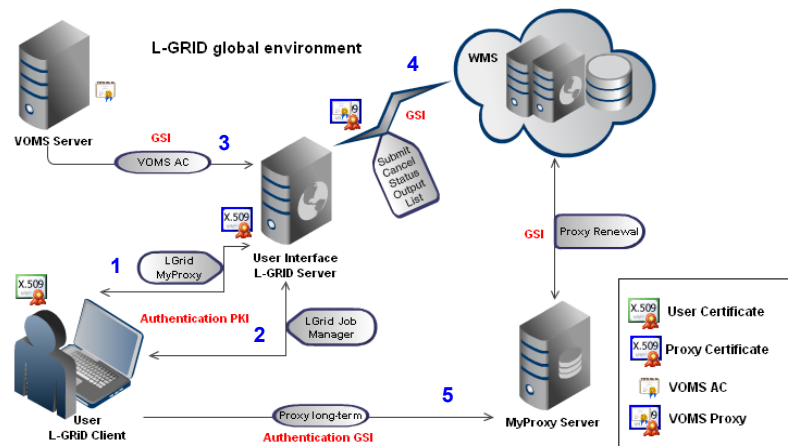

Figure 1: L-GRID architecture

The Figure 1 shows the procedure used in order to authenticate a user, and the user's interaction with the EGI Grid infrastructure:
1. authentication of a user requiring access to the portal using a web browser on his pc, and delegation of the user's credential to the L-GRID server;
2. interaction with the L-GRID server, with respect to the job submission and management;
3. data exchange between the server and the VOMS (Virtual Organization Membership Service) for the user authorization request;
4. communication with the WMS (Workload Management System, responsible for the central services Grid management);
5. the system used by EGI in order to ensure a Proxy certificate renewal based on the MyProxy service.

The novel idea introduced by L-GRID has been the split into client and server of the main gLite User Interface commands [5]. This entails an increased security level in Proxy Certificates management: it is no more necessary to send username and password to the MyProxy server for delegate a proxy certificate or have a copy of the X.509 personal certificate on the User Interface.

At the same time a smallest amount of data (applications and certificates) needs to be transmitted over the network: the job input and output files are automatically compressed.

**4. Dynamic credentials delegation**

Through the implementation of a mechanism for dynamic delegation on the server side of the portal, we extend the concept of Grid to the connected clients too. The user client becomes itself part of the Grid infrastructure.

The end user needs a valid X.509 personal certificate [6] [7], and an access to the Internet. The X.509 personal certificate does not get out from the local machine, strictly compliant to the security policies.

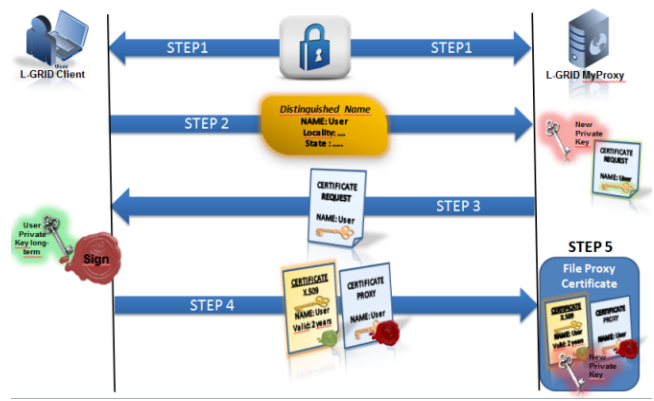

Figure 2: Dynamic credentials delegation

When a user is trying to communicate to the L-GRID server in order to create a Proxy certificate based on his X.509 personal certificate, the invoked procedure needs to be fully compliant to the policies imposed by the Certification Authorities and by the certificates on Grid rules.

The solution comes directly from the Kesselman, Foster et al. paper *"X.509 Proxy Certificates for Dynamic Delegation"* [8]; it consists of a dynamic delegation mechanism to create a Proxy certificate on the server side (Figure 2):
1. over a secure SSL communication, the client sends the user Distinguished Name to the server;
2. the server generates a key pair (public and private) and a Certificate Signing Request (composed of the client DN, and the public key just generated);
3. the server sends the Certificate Signing Request to the client;
4. the client signs the Certificate Signing Request with the user personal X.509 private key and sends to the server the Proxy certificate just created and its X.509 public certificate;
5. the server is now able to create a file Proxy certificate, composed of: the client X.509 public certificate and the Proxy certificate sent by the client, with the private key generated by the server.

In L-GRID the user ID is derived from the Distinguished Name (DN) of the X.509 personal certificate. The main differences with respect to a native User Interface are the extreme ease of use and the no-need of users registration. This way the end user needs only her/his personal X.509 certificate, issued from a Certification Authority, and an access to the Internet.

In our case, the MyProxy server is contacted only to renew the proxy in long term jobs, if required. It allows to reduce the time spent for the job submission, ensuring at the same time a higher efficiency and a better security level in proxy delegation and management.

## 5. Requirements and features

### 5.1.1 Requirements

No users registration is required in order to use the L-GRID portal. The user is authenticated by her/his X.509 personal certificate, issued from a trusted Certification Authority. This way there is no needs of a pre-created user account over the central servers managing the whole infrastructure.

### 5.1.2 Features

The L-GRID gateway provides a full control over the complete lifecycle of a Grid Job: certificate conversion, job submission, status monitoring, output retrieval. It provides also a very simple and customizable JDL editor.

On the client side, at the first use of the portal, a 3 MB Java applet is downloaded, in order to manage the X.509 proxy certificate delegation and the communications security.

The system is user-friendly, secure - it uses SSL protocol, mechanism for dynamic delegation and identity creation in public key infrastructures - highly customizable, open source, and easy to install; the package setup requires a few MB.

### 5.1.3 The use of the L-GRID web portal

Accessing the portal via web, at the first access the user is required to convert his X.509 personal certificate from .p12 to .pem format. This is done by the portal.

In Figure 3 is shown the simple JDL editor (on the left), allowing the user to create his job. On the right figure is shown the Submit tab. Using this graphic interface the user may choose the JDL, the directory containing his working files, the proxy lifetime, and the MyProxy server for the proxy certificate renewal - if required.

As the final step, the user needs to select the Virtual Organization he is member of, and insert his X.509 personal certificate password. The Submit button starts the submission procedure.

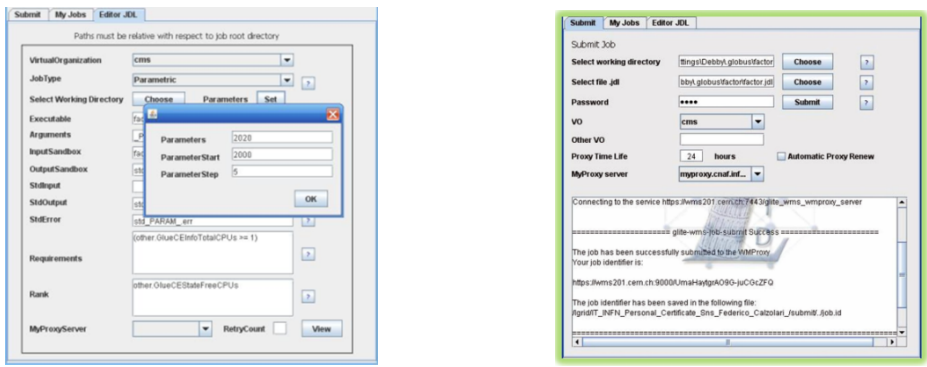

Figure 3: JDL editor, Job Submit

On the My Jobs tab (Figure 4), the user is able to see the status of his jobs - running, terminated, pending - and attend to the output retrieval of the finished jobs. Different colors are used to tag the several possible status of a job: green for successfully done, red for aborted, and other for cleared, submitted, running, ready.

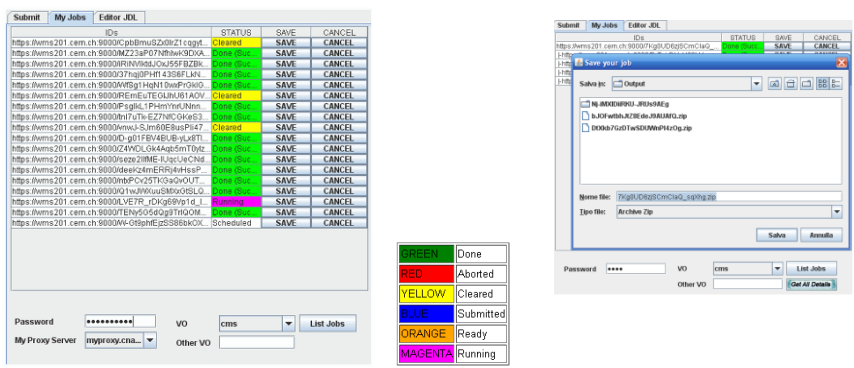

Figure 4: Job monitoring, Output retrieval

By simply selecting a job, the user is able to retrieve its output and save it in the desired directory, or clear the job by sending a job terminate to the Workload Management System.

With respect to the parametric jobs or the jobs collections, L-GRID offers the possibility to manage independently both the whole computation process, and the single jobs composing the collection.

## 6. Tests and performances

The integration of the dynamic credentials delegations inside the client - server architecture, avoiding the use of a third part MyProxy server, allows to reduce the time spent for the job submission, monitoring, output retrieval, providing a higher efficiency.

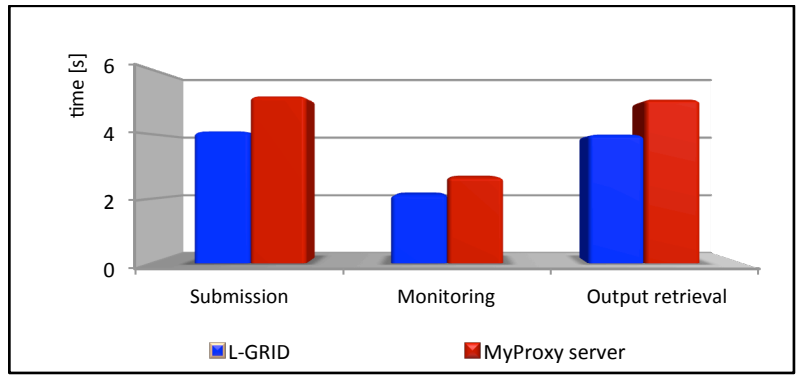

Figure 5: L-GRID vs MyProxy server performances

This entails an improvement in the efficiency, as shown in Figure 5: in blue the time in seconds required by L-GRID job submission, monitoring, and output retrieval; in red the same operations are executed using a MyProxy external server.

## 7. Security

An extra security improvement has been achieved implementing on site a dynamic delegation mechanism, according to the Grid Security Infrastructure [9]. It allows to guarantee a higher efficiency and a better security level in proxy delegation and management.

In order to manage the proxy certificate delegation, the dynamic delegation service inside L-GRID portal avoids the need of sending username and password to the MyProxy server for delegating a proxy certificate. This entails an increased security level in proxy certificates management, by avoiding the possibility for a malicious attack to retrieve the Proxy certificates stored with password in the MyProxy server.

## 8. Conclusions and future work

The L-GRID gateway [10] [11], developed at the Scuola Normale Superiore in collaboration with the University of Pisa [Italy] and the Italian National Institute for Nuclear Research INFN, is intended to be a helpful tool to access Grid resources shared all around the world via a simple Web interface, using whatever operating system and browser, with no registration required at all.

By the use of distributed Web portals, viewed as a part of the computing facilities, integrated in a Grid computing infrastructure, many user communities could be able to expand their computational power, in order to speed up the results of their research.

The portal has been chosen by the Theophys Virtual Organization as login and job submission facility for its users.

The results obtained encourage future developments. Further steps are represented by the integration with a MyProxy server locally hosted for long term job management.